\documentclass[10pt]{sigplanconf}

\pdfoutput=1

\usepackage[T1]{fontenc} 
\usepackage[utf8]{inputenc} 
\usepackage{graphicx}
\usepackage{ifthen}
\usepackage{xspace}
\usepackage{alltt}
\usepackage{latexsym}
\usepackage{url}            
\usepackage{amssymb}
\usepackage{amsfonts}
\usepackage{amsmath}
\usepackage{stmaryrd}
\usepackage{enumerate}
\usepackage{xspace}
\usepackage{tikz}
\usepackage{fancybox}

\newboolean{showcomments}
\setboolean{showcomments}{false}

\ifthenelse{\boolean{showcomments}}
  {\newcommand{\bnote}[2]{
	\fbox{\bfseries\sffamily\scriptsize#1}
    {\sf\small$\blacktriangleright$\textit{#2}$\blacktriangleleft$}
   }
   
  }
  {\newcommand{\bnote}[2]{}
   
  }

\newcommand{\tincho}[1]{\bnote{Martin}{#1}}

\graphicspath{{figures/}}

\newcommand{\commented}[1]{}

\newcommand{\stSelector}{$\gg$}

\newcommand{\eg}{\emph{e.g.,}\xspace}
\newcommand{\ie}{\emph{i.e.,}\xspace}
\newcommand{\etal}{\emph{et al.}\xspace}
\newcommand{\ct}[1]{{\textsf{#1}}\xspace}

\newcounter{enodecounter}
\newcounter{elastnodecounter}

\newenvironment{code}
    {\begin{alltt}\sffamily\small}
    {\end{alltt}\normalsize}

\usepackage{url}            
\makeatletter
\def\url@leostyle{%
  \@ifundefined{selectfont}{\def\UrlFont{\sf}}{\def\UrlFont{\small\sffamily}}}
\makeatother
\newtheorem{definition}{Definition}

\begin{document}

\newcommand{\Approach}{\textsc{DeltaImpactFinder}\xspace}
\newcommand{\Problem}{semantic merge conflict\xspace}
\newcommand{\Problems}{\Problem{}s\xspace}

\conferenceinfo{IWST'15}{July 15 - 16, 2015, Brescia, Italy}
\copyrightyear{2015} 
\copyrightdata{978-1-4503-3857-8/15/07} 
\doi{2811237.2811299}
\exclusivelicense                

\title{\Approach: Assessing Semantic Merge Conflicts\\with Dependency Analysis}

\authorinfo{Mart\'in Dias \and Guillermo Polito \and Damien Cassou \and St\'ephane Ducasse}{
RMoD\\
Inria Lille--Nord Europe --- University of Lille --- CRIStAL, France}

\maketitle

\begin{abstract}

In software development, version control systems (VCS) provide branching and merging support tools. 
Such tools are popular among developers to concurrently change a codebase in separate lines and reconcile their changes automatically afterwards.
However, two changes that are correct independently can introduce bugs when merged together.
We call \emph{\Problems{}} this kind of bugs.

Change impact analysis (CIA) aims at estimating the effects of a change in a codebase. 
In this paper, we propose to detect \Problems{} using CIA.
On a merge, \Approach analyzes and compares the impact of a change in its origin and destination branches.
We call the difference between these two impacts the \emph{delta-impact}.
If the delta-impact is empty, then there is no indicator of a \Problem{} and the merge can continue automatically.
Otherwise, the delta-impact contains what are the sources of possible conflicts.



\end{abstract}

\section{Introduction}

Software projects are in constant evolution.
Often, developers perform changes concurrently in a codebase, generating separate lines of development.
Version control systems (VCS) support this activity through branches, a widely used feature~\cite{Dias14b, Phil11a} in software development. 
Merge~\cite{Mens02d} (also called integration) is a fundamental operation in VCS that reconciles two (or more) branches. 
VCS can detect syntactical merge conflicts automatically.
Nevertheless, \Problems (\ie at the level of program behavior) exceed the scope of these tools.
Consider, for example, a branch renaming a template method from \ct{A>>foo} to \ct{A>>bar} and another branch overriding \ct{foo} in a \ct{B} class, subclass of \ct{A}. 
The two branches can be automatically merged but the resulting code will fail to execute as intended: indeed \ct{B>>foo} will never be executed while \ct{B>>bar} is supposed to exist but does not,
Such \emph{\Problems} are not detected by current VCS. 




Change Impact Analysis~\cite{Bohn96a, Li12} (CIA) is an active research field that aims at identifying the potential consequences of a change in a codebase.
Typically, CIA techniques establish \emph{dependency} relationships between the code entities of the codebase.
These relationships are used afterwards for detecting the set of entities that are impacted by a change.
The rationale is that when a code entity changes, the behavior of the \emph{dependent} entities is impacted.
Many CIA research works use the technique of computing the dependencies of a change in the original codebase where the authors created the change \cite{Petr09a, Canf05b, Hass06a, Anto00b}.



This paper proposes a solution to help integrators in the detection of \Problems{} using CIA.
On a merge, \Approach analyzes and compares the impact of a change in its origin and destination branches.
We call the difference between these two impacts the \emph{delta-impact}.
If the delta-impact is empty, it means that there is no \Problem{} and the merge can continue automatically.
Otherwise, the delta-impact contains what are the sources of possible conflicts.
The contributions of this paper are the following:

\begin{itemize}
\item a description of \Problem with an example (Section \ref{sec:problem});
\item a CIA technique, named \Approach, to detect \Problems (Section \ref{sec:solution});
\item a discussion of usages of this technique (Section \ref{sec:discussion});
\item a prototype of this technique implemented in Pharo (Section \ref{sec:implementation});
\end{itemize}

\newpage
\section{Problems when Merging: Semantic Conflicts} \label{sec:problem}

To show how semantic conflicts appear when merging, we start by introducing an example of the Fragile Base Class Problem~\cite{Mikh98a} (FBCP).
Consider the following logging library that implements a \ct{Log} class whose API has the methods \ct{log:}, which can record a single message into an internal collection, and \ct{logAll:}, which records multiple messages in one shot using \ct{log:}. 
The logic for adding an element to the collection of logs is only expressed inside the \ct{log:} method.
Following there is the code illustrating the most relevant points of such implementation:

\begin{code}
Object subclass: #Log
   instanceVariableNames: 'messages'.

Log \stSelector log: aMessage  
   messages add: aMessage.

Log \stSelector logAll: someMessages  
   someMessages do: [:each | self log: each ].
\end{code}

\begin{figure}[ht]
\begin{center}
\includegraphics[width=0.95\columnwidth]{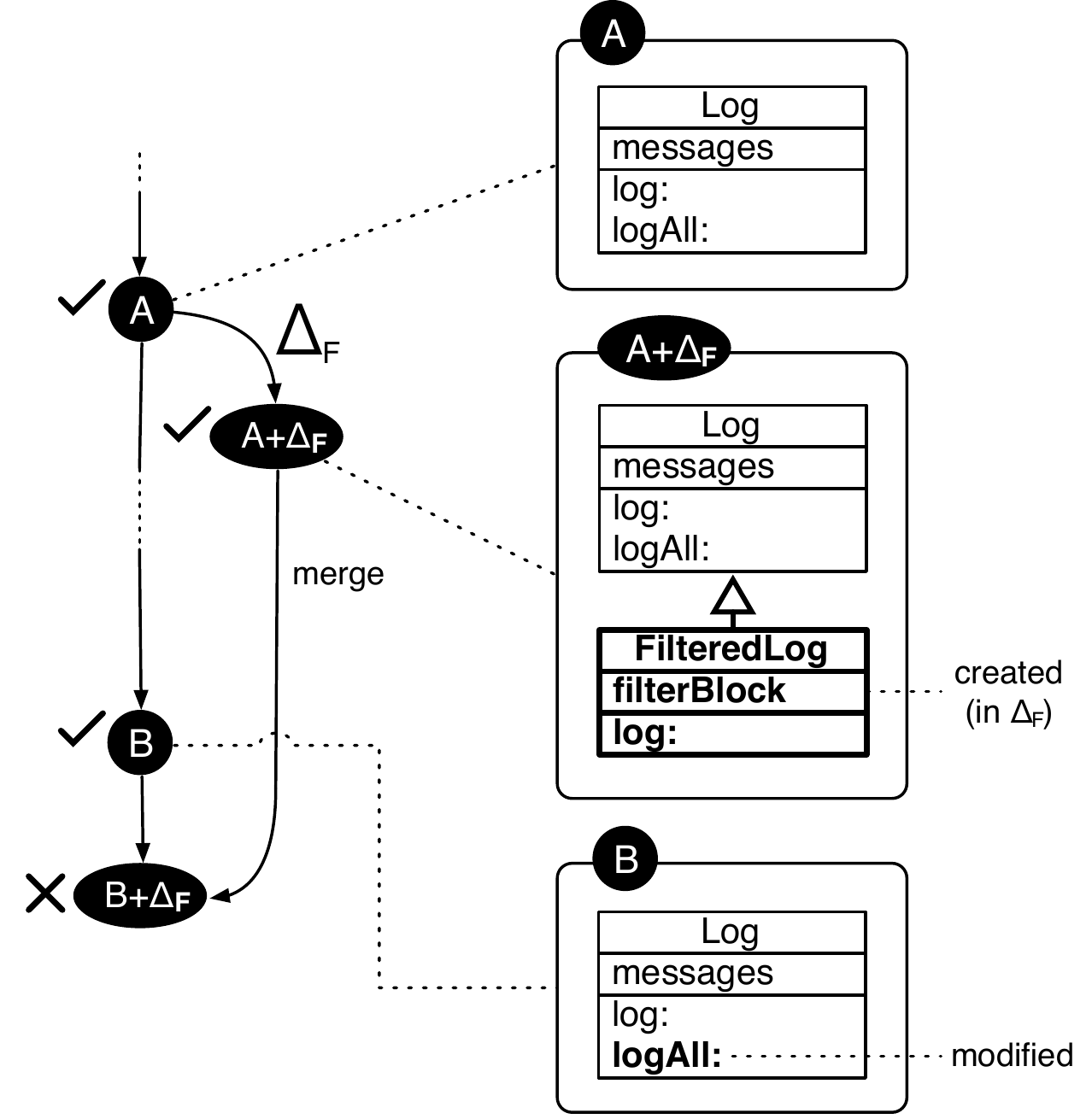}
\bigskip
\caption{
A developer starts a new branch of the library from version $A$ of a logging library and implements a new feature: the \ct{FilteredLog}. We name such change as $\Delta_{F}$. Before $\Delta_{F}$ is integrated into the library, a modification in the method \ct{Log\stSelector logAll:} in $B$ makes $B+\Delta_{F}$ not working.}
\label{exampleMerging}
\end{center}
\end{figure}

We want now to introduce a change in this library.
At some point in time, a developer starts a new branch of the library from version $A$ and implements a new feature: the \ct{FilteredLog}~(Figure~\ref{exampleMerging}). \ct{FilteredLog} is a subclass of \ct{Log} that overrides \ct{log:} to record the message only when it satisfies a filter. 
Note that \ct{FilteredLog} does not need to override \ct{Log\stSelector logAll:}.
We refer to this change as $\Delta_{F}$.
The code illustrating $\Delta_{F}$ is the following:

\begin{code}
\texttt{+} Log subclass: #FilteredLog
\texttt{+}    instanceVariableNames: 'filterBlock'.
\texttt{+}
\texttt{+} FilteredLog \stSelector log: aMessage  
\texttt{+}    (filterBlock value: aMessage) 
\texttt{+}       ifTrue: [ super log: aMessage ].
\end{code}

In parallel, the main branch of the library evolves: the method \ct{Log\stSelector logAll:} no longer uses \ct{self log:} to record each received message, but instead each message is added directly to the internal collection.

\begin{code}
\texttt{ } Log \stSelector logAll: someMessages  
\texttt{-}    someMessages do: [:each | self log: each ].
\texttt{+}    messages addAll: someMessages
\end{code}

When the integrator wants to merge $\Delta_{F}$ in the main branch, the tool does not inform any merging conflicts but the introduced feature does not work as expected. Indeed, \ct{FilteredLog} does not filter any messages when using \ct{logAll:} because this method does not use the \ct{log:} message anymore.

\begin{code}
log := FilteredLog new.
log filterBlock: [ :each | each > 0 ].
log logAll: \#(-5).
log messages isEmpty. "false ---> wrong!"
\end{code}

We can observe from this example that an integrator can merge $\Delta_{F}$ introducing a semantic conflict silently.
To detect such kind of bugs, integrators need to understand the code in a change ($\Delta$) more deeply \eg know the intention of the change, in which version was it developed.
Then, the activity of integration requires a big human effort which new tools can help to alleviate.

In a more general way, we would like a tool that helps the integrator by answering the following questions:

\begin{description}
\item[Q1.] \emph{Does a $\Delta$ produce semantic conflicts if merged in a particular version of the codebase?}
\item[Q2.] \emph{What code entities are involved in a semantic conflict that $\Delta$ produces?}
\item[Q3.] \emph{When was the change that produced a semantic conflict with a $\Delta$ integrated?}
\end{description}

In general these questions are hard to answer, specially in large scale projects where developers work in parallel on the same codebase. 
Automated Testing and Continuous Integration practices could help in answering the first question. 
However, these practices depend on the coverage of the testing: the more tested is the code, the more likely the problem can be detected. 
Unfortunately, sometimes test coverage is not good enough and requires an effort that developers do not make.
Then, we pose the following research question:

\begin{center}\emph{Can CIA on origin and destination branches help to answer these questions?}\end{center}

\section{Our Solution: \Approach} \label{sec:solution}

%

%
%
%
%
%
%
%
%
%
%
%
%
%
%
%
%

In a nutshell, we propose to detect semantic merge conflicts using CIA.
On a merge, \Approach analyzes and compares the impact of a change in its origin and destination branches.
We call the difference between these two impacts the \emph{delta-impact}.
If the delta-impact is empty, it means that there are no semantic conflicts and the merge can continue automatically.
Otherwise, the delta-impact contains what are the sources of possible conflicts.

In the following we describe our approach. 
First, we define the notions of dependency and impact. 
Then, based in such notions, we explain delta-impact.

\subsection{Dependency and Impact}

Change Impact Analysis~\cite{Bohn96a, Li12} (CIA) aims at identifying the potential consequences of a change in a codebase.
Typically, CIA techniques establish \emph{dependency} relationships between the code entities of the codebase, which they use afterwards for detecting the set of entities that are impacted by a change.
The rationale is that when a code entity changes, the behavior of the \emph{dependent} entities is impacted.
In the context of this paper, we define dependency as follows:

\begin{definition}{Dependency.}
A \emph{dependency} is the relationship between two code entities where one code entity ($source$) requires the other ($target$). We denote it $source \to target$.
\end{definition}

In the motivational example we introduced in Section \ref{sec:problem}, the \ct{FilteredLog} class depends on the \ct{Log} class because of the inheritance relationship between them.
In this paper we focus on static dependency analysis, \ie dependencies that are explicit in the source code, however we believe that \Approach can be generalized to other kinds of dependencies. We describe the dependencies of \Approach in more detail in Section~\ref{subsec:dependencies}.

Then, we define the impact of a $\Delta$ as follows:

\begin{definition}{Impact.}
The \emph{impact} of a change $\Delta$ in a codebase $C$, denoted $I(\Delta, C)$, is the set of dependencies introduced or removed in $C$ after applying $\Delta$. 
\end{definition}

\begin{figure}[ht]
\begin{center}
\includegraphics[width=0.72\columnwidth]{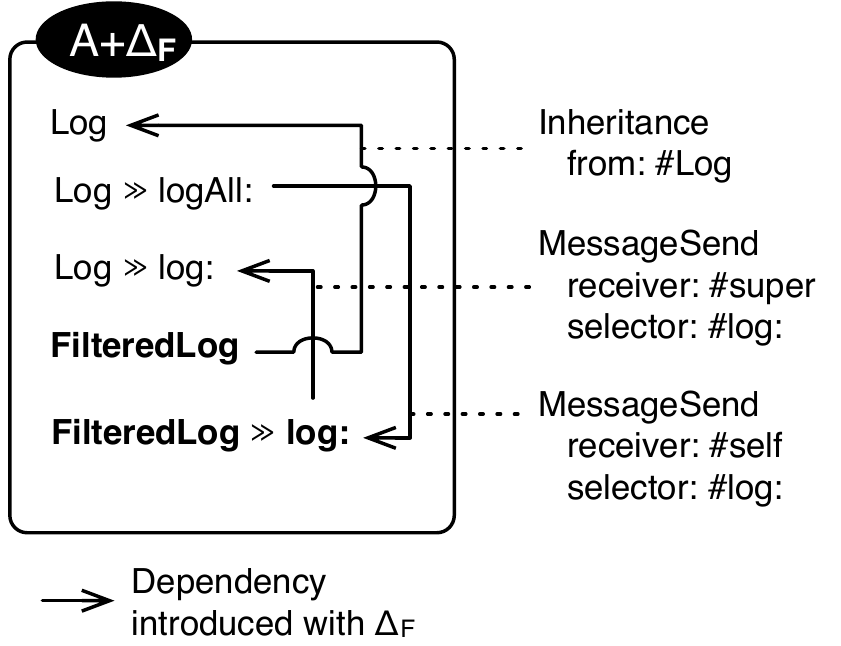}
\smallskip
\caption{Impact of $\Delta_{F}$ in $A$. In the origin branch, $\Delta_{F}$ introduces three dependencies: one corresponding to an inheritance relationship, and the others corresponding to message sends.}
\label{exampleMergingDependencies}
\end{center}
\end{figure}

In the motivational example, the impact of $\Delta$ in its origin branch includes the following dependency modifications (Figure~\ref{exampleMergingDependencies}):

\begin{center}
\begin{minipage}[t]{0.85\columnwidth}
\begin{description}
\item [$i_1$] Introduction of an inheritance dependency from \ct{FilteredLog} to \ct{Log}.
\item [$i_2$] Introduction of a message send dependency from \ct{Log>>logAll:} to \ct{FilteredLog>>log:}.
\item [$i_3$] Introduction of a message send dependency from \ct{FilteredLog>>log:} to \ct{Log>>log:}.
\end{description}
\end{minipage}
\end{center}

\subsection{\Approach}

\begin{figure}[ht]
\begin{center}
\smallskip
\includegraphics[width=0.68\columnwidth]{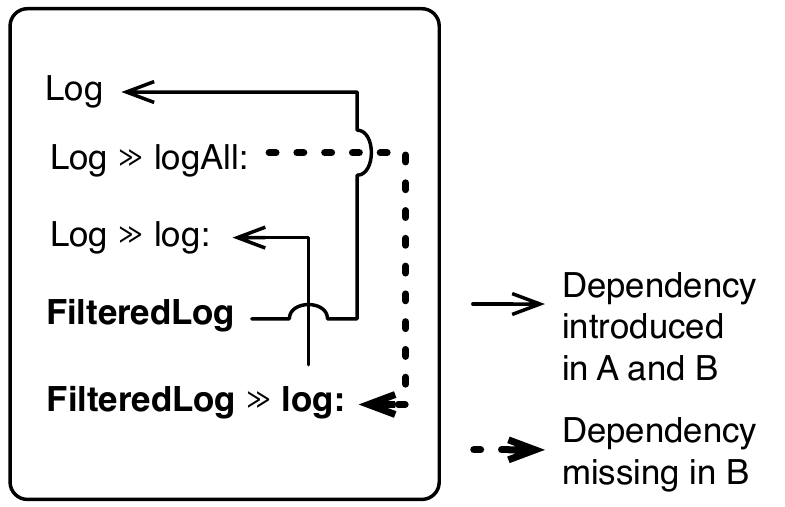}
\smallskip
\caption{\textbf{Delta-impact of $\Delta_{F}$ from $A$ to $B$.} Comparison of the impact of $\Delta_{F}$ in both $A$ and $B$. The dependency from \ct{Log>>logAll:} to \ct{FilteredLog>>log:} is missing in $B$.}
\label{exampleDependencies}
\end{center}
\end{figure}

In the example~(Figure~\ref{exampleDependencies}), the comparison of the impact of $\Delta_{F}$ in origin and destination branches shows that the dependency from \ct{Log>>logAll:} to \ct{FilteredLog>>log:} is missing in the destination branch.
Precisely, the cause of the semantic conflict in the example is the change in \ct{Log>>logAll:}, which no longer invokes \ct{FilteredLog>>log:}.

We observe that the impact of a $\Delta$ contains a set of relationships between code entities in a particular version of code.
Informally, we can think about this impact as a set of constraints that have to be satisfied for the code to work as expected.
Then, we pose the following hypothesis:

\bigskip
\begin{minipage}[t]{0.95\columnwidth}
\begin{center}\emph{A semantic conflict appears when the set of dependencies that a $\Delta$ introduces (or removes) into its origin branch is different from those introduced (or removed) when merging such $\Delta$ in the destination branch.}\end{center}
\end{minipage}
\bigskip

In other words, if there are missing or extra dependencies in the destination branch, this could mean that $\Delta$ has a different meaning than the one intended by its author. 
On the contrary, if the dependencies are the same the change may have the same effects.
Then, we define \emph{delta-impact} as follows:



\begin{definition}{Delta-Impact.}
The \emph{delta-impact} of a change $\Delta$ with origin branch $A$ and destination branch $B$, denoted $DI(\Delta, A, B)$, is the symmetric difference\footnote{the symmetric difference between two sets includes only the elements that belong to one of such sets but not to both.} between the set of impacts of $\Delta$ in $A$ and the impact of $\Delta$ in $B$.
\end{definition}

To compute the delta-impact of $\Delta$, we obtain the impact of $\Delta$ in the origin and destination branches, and then we compute the symmetric difference between the two impacts. 
In the example (Figure~\ref{exampleDeltaImpact}), the impact $I(\Delta_{F}, A)$ yields the set $\{i_1, i_2, i_3\}$, while $I(\Delta_{F}, B)$ yields the set $\{i_1, i_3\}$.
Then, $DI(\Delta_{F}, A, B)$ results in the set $\{i_2\}$, because $i_2$ is missing in the destination branch $B$.
In this context, a tool that computes and shows the delta-impact of $\Delta_{F}$ to the integrator would answer the questions Q1 and Q2 that we defined in Section~\ref{sec:problem}.

\begin{figure}[ht]
\small

\[
\begin{array}{ccrclc}
\multicolumn{5}{l}{I(\Delta_{F}, A):}\\
& + & \ct{FilteredLog} & \rightarrow & \ct{Log} & (i_1) \\
& + & \ct{Log>>logAll:} & \rightarrow & \ct{FilteredLog>>log:} & (i_2) \\
& + & \ct{FilteredLog>>log:} & \rightarrow & \ct{Log>>log:} & (i_3) \\
\\
\multicolumn{5}{l}{I(\Delta_{F}, B):}\\
& + & \ct{FilteredLog} & \rightarrow & \ct{Log} & (i_1) \\
& + & \ct{FilteredLog>>log:} & \rightarrow & \ct{Log>>log:} & (i_3) \\
\\
\multicolumn{5}{l}{DI(\Delta_{F}, A, B) = I(\Delta_{F}, A) \ominus I(\Delta_{F}, B):}\\
& + & \ct{Log>>logAll:} & \rightarrow & \ct{FilteredLog>>log:} & (i_2) \\
\end{array}
\]

\caption{\textbf{\Approach} $I(\Delta_{F}, A)$ is the original impact of $\Delta_{F}$, while $I(\Delta_{F}, B)$ is the destination impact. $DI(\Delta_{F}, A, B)$ is the delta-impact of $\Delta_{F}$ in the destination branch, which shows that a dependency introduction is missing ($i_2$).}
\label{exampleDeltaImpact}
\end{figure}

\section{Applicability}
\label{sec:discussion}

In this section we describe concrete scenarios where we would like to validate \Approach (in future works).
Since our plan is to validate our approach with Pharo community,  
we present these scenarios in terms of Pharo.

\subsection{Aged Code Changes}\label{subsec:aged}

The codebase of a large open-source project like Pharo evolves through code changes that developers submit.
When a developer submits a code change, this change must pass a reviewing process before an integrator merges it into the main development branch.
Since this reviewing process takes some time and code changes are integrated every day, when the integrator has to merge an approved code change, the current Pharo codebase may be different than the codebase where the author of the change worked.

In this situation, \Approach can help Pharo integrators to discover semantic conflicts on the merge.
Given a code change $\Delta$ submitted by a developer, \Approach can answer:
\begin{center}
``Is the impact of $\Delta$ in the current Pharo the same as the impact in the Pharo where $\Delta$ was originally created?''
\end{center}


\subsection{Software Migration}\label{subsec:migration}

Often, when the codebase of a project changes, other projects that depend on it need to be migrated.
This change propagation is known as ripple effect~\cite{Yau78}.
Ripple effects are problematic because a small change in a project can have a very large impact on other projects.
Additionally, sometimes a code that needs migration remains undiscovered for a long time due to low test coverage.

We can illustrate this scenario by rephrasing the FBCP example used in Section~\ref{sec:problem}. 
Let's suppose that a developer works in a project in Pharo version $A$.
The system provides the class \ct{Log}, which the developer extends in his project by creating the subclass \ct{FilteredLog}.
One year later, a new stable version of Pharo is available: version $B$. 
Among plenty of changes in Pharo $B$, the method \ct{Log>>logAll:} has been modified like in the FBCP example. 
As before, a bug appears in the method \ct{logAll:} when invoked in an instance of \ct{FilteredLog}.
Note that when the developer loads the \ct{FilteredLog} class in Pharo $B$, the system does not raise any load or compilation error: it is another form of the \Problem.
The developer will probably have to debug his project to find that the change in \ct{Log>>logAll:} is the responsible.
If a tool would have informed the developer that \ct{Log>>logAll:} changed, then he could save time.

We can pose this problem in terms of our approach. 
When the \ct{FilteredLog} package is loaded in Pharo $A$, some dependencies are introduced between code entities of \ct{FilteredLog} and Pharo.
This is what we have defined as impact.
When the package is loaded in Pharo $B$, the impact is different: $i_{2}$ is missing (Figure~\ref{exampleDeltaImpact}).
In general terms, \Approach can help developers to answer the following question:
\begin{center}
``Which Pharo code entities with impact on my project changed from Pharo $A$ to $B$?''
\end{center}

%

\subsection{Requirements} \label{subsec:requirements}

The main requirement of any CIA technique is a high precision and a high recall~\cite{Li12}. 
A high precision means that a technique finds substantially more relevant results than irrelevant, while high recall means that a technique finds most of the relevant results.

However, we extract some additional requirements for the implementation of \Approach from the scenarios presented above in this section:

\begin{description}
\item[Isolation from tool's environment.]
For supporting ``Software Migration'', the implementation needs to compute dependencies of code entities as if they were loaded in some arbitrary Pharo version, independently of the Pharo version where the tool is actually running.


\item[Usable in real use cases.] Since we aim at building a tool that real developers can evaluate, the implementation should compute the dependencies in a reasonable time.

\end{description}

\section{Implementation}
\label{sec:implementation}

We start this section doing an overview of the main characteristics of our prototype implementation of \Approach.
Some design decisions are consequence of the requirements stated in Section~\ref{subsec:requirements}.
\begin{description}

\item[Static code analysis.] 
We use default Pharo support for performing static code analysis.
For example, the \ct{AST-Core} package provides support for visiting the \emph{abstract syntax tree} of methods and collecting dependencies.

\item[Light-weight and polymorphic code metamodel.]
We implemented \ct{RingFicus}, a metamodel for Pharo code entities.
It provides first-class representations for class, method, instance variable, etc.
RingFicus allows to model code either internal or external to the current Pharo environment to browse them, query them, analyze them, as if they were loaded in the system.




\end{description}






\subsection{Computing Dependencies}\label{subsec:dependencies}

\Approach requires to compute dependencies between source code entities. In our solution such entities are classes, metaclasses, instance and class variables, traits, class-traits and methods.
The relationships that we consider as dependencies in \Approach are the following:

\begin{description}
\item[Inheritance:] A dependency from a class to its superclass.
\item[Trait Usage:] A dependency from a class, metaclass, trait, or class-trait to all traits in its \emph{trait composition}.
\item[Variable Access:] A dependency from a method in a class or metaclass that accesses (read or write) an instance or class variable, to the accessed variable. 
\item[Message Send:] A dependency from a method including a message-send sentence, to all the possible methods that are invoked. 
Due to the absence of type information in the language, in the general case the algorithm uses the selector of a message-send to look up for all the implementors in the codebase.
However, in the case of \ct{self}-sends and \ct{super}-sends the algorithm refines its look up for obtaining more precise dependencies.
\end{description}

For testing our prototype, we implemented \ct{DependencyMiner}, which iterates over all the source code entities of a Pharo environment collecting the dependencies.
Each dependency is an association $source \to target$.

\subsection{Computing Impact and Delta-Impact}

\begin{figure}[ht]
\begin{center}
\includegraphics[width=0.5\columnwidth]{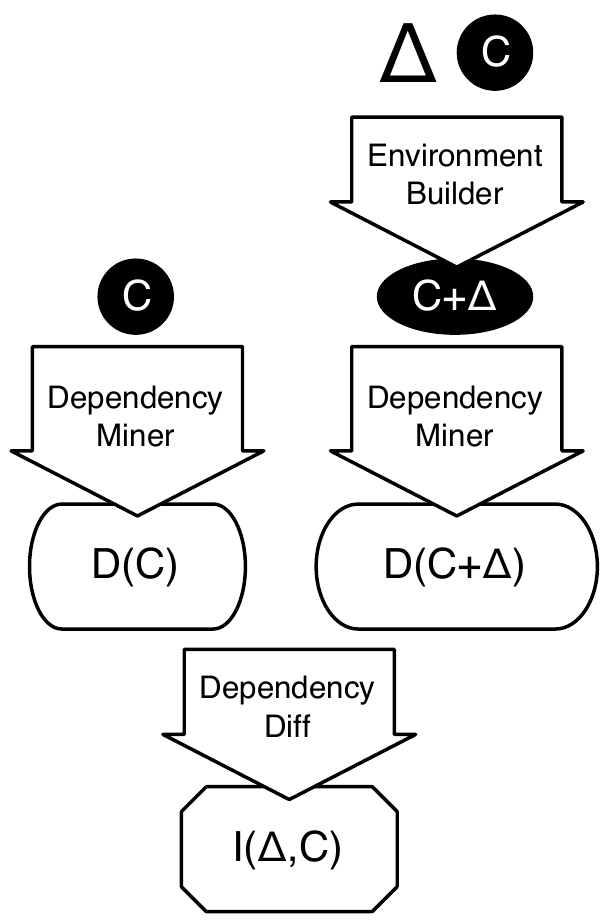}
\bigskip
\caption{\textbf{Computation of the impact of $\Delta$ in $C$.} The algorithm starts by building the codebase $C+\Delta$. After, the algorithm computes the dependency sets of each codebase ($D(C)$ and $D(C+\Delta)$). Finally, the algorithm computes the impact by finding the symmetric difference between $D(E)$ and $D(E+\Delta)$.}
\label{ImpactBuildOverview}
\end{center}
\end{figure}

For the computation of the impact of $\Delta$ in an environment (Figure~\ref{ImpactBuildOverview}), the algorithm starts by building the codebase $C+\Delta$.
Then, the prototype computes the dependency sets of each environment using the \ct{DependencyMiner}, described above.
Finally, the prototype the impact and the delta-impact by performing \ct{Collection} operations.


%

%
%
%
%

%
%

\section{Related Work}

\paragraph{Change Impact Analysis.}
In an exhaustive survey~\cite{Li12} about CIA, Li \etal analyze 30 publications from 1997 to 2010 and identified 23 code-based CIA techniques. 
The study characterizes the CIA techniques, and identifies key applications of CIA techniques in software maintenance.
Typically, CIA techniques establish \emph{dependency} relationships between the code entities of the codebase, which they use afterwards for detecting the set of entities that are impacted by a change.
The rationale is that when a code entity changes, the behavior of the \emph{dependent} entities is impacted. 
The techniques to identify dependencies in a codebase are typically classified into static and dynamic.
The static techniques identify the dependencies using static code analysis~\cite{Wels84a, Sun10a}, while dynamic techniques~\cite{Kore98a, Apiw05a} collect data from program execution.
There are, as well, mixed techniques~\cite{Cai15a} which combine both techniques.
\Approach is orthogonal to the technique to identify dependencies, besides our prototype works with a static CIA technique. 


\paragraph{Merging.}
The \Problems have been studied before under different names.
Mens~\cite{Mens02d} describes this problem in his survey of code merging.
In this work, the author remarks that most approaches to software merging have been validated on imperative programming languages and it is not trivial to port these approaches to the object-oriented paradigm, due to late binding and polymorphism in object-oriented programming languages.
\tincho{add more refs -> Good references in Veronica's PhD}
\tincho{link with our approach}

\paragraph{Ring Metamodel.}
Ring~\cite{Uqui11a}\cite{Uqui10b} is a source code meta-model that serves as a unified infrastructure for building tools in Pharo.
While Ring has proven efficacy for dependency analysis tools~\cite{Uqui14a, Uqui12b}, we found some limitations that driven us to reimplement our own RingFicus package.
In our early tests, Ring did not fulfill the requirements we described in Section~\ref{sec:discussion}.
Ring code entities did not ensure isolation from tool's environment, and they were not efficient to represent a whole Pharo environment.

\section{Conclusion and Future Perspectives}

In this paper, we proposed a solution to help integrators in the detection of \Problems{} using CIA.
On a merge, \Approach analyzes and compares the impact of a change in its origin and destination branches.
We call the difference between these two impacts the \emph{delta-impact}.
If the delta-impact is empty, it means that there is no \Problem{} and the merge can continue automatically.
Otherwise, the delta-impact contains what are the sources of possible conflicts.

In short, this paper makes the following contributions:
\begin{itemize}
\item a description of \Problem with an example;
\item a CIA technique, named \Approach, to detect \Problems; \tincho{novel?}
\item a discussion of concrete scenarios to validate this technique in future work;
\item a prototype of this technique implemented in Pharo.
\end{itemize}

\section*{Acknowledgements} We would like to thank Guille Polito, Santiago Bragagnolo and Lucas Godoy for their support during this work. This work was supported by Ministry of Higher Education and Research, Nord-Pas de Calais Regional Council.

\bibliographystyle{plain}
\bibliography{rmod,others}

\end{document}